# Charge state distribution analysis of Al and Pb ions from the laser ion source at IMP*

SHA Shan(沙杉) [1,2,3;1)]  JIN Qian-Yu(金钱玉)[3]  LI Zhang-Min(李章民)[3]  GUO Xiao-Hong(郭晓虹)[2]  Zhou Lun-Cai(周伦才)[2]  CAI Guo-Zhu(蔡国柱)[2]  SUN Liang-ting(孙良亭)[2]  Zhang Xue-Zhen(张雪珍)[2]  ZHAO Huan-Yu(赵环昱)[2]  CHEN Xi-Meng(陈熙萌)[1]  ZHAO Hong-Wei(赵红卫)[2]

[1] School of Nuclear Science and Technology, Lanzhou University, Lanzhou 730000, China

[2] Institute of Modern Physics, Chinese Academy of Sciences, Lanzhou 730000, China

[3] University of Chinese Academy of Sciences, Beijing 100049, China

**Abstract:** A prototype laser ion source that could demonstrate the possibility of producing intense pulsed high charge state ion beams has been established with a commercial Nd:YAG laser ($E_{max} = 3$, 1064 nm, 8-10 ns) to produce laser plasma for the research of Laser Ion Source (LIS). At the laser ion source test bench, high purity (99.998 %) aluminum and lead targets have been tested for laser plasma experiments. An Electrostatic Ion Analyzer (EIA) and Electron Multiply Tube (EMT) detector were used to analyze the charge state and energy distribution of the ions produced by the laser ion source. The maximum charge states of $Al^{11+}$ and $Pb^{19+}$ were achieved. The results will be presented and discussed in this paper.

*Supported by China Nature Science Foundation (10921504, 11275239)

1) E-mail: s.shan@impcas.ac.cn

# 1 Introduction

Many kinds of high current multiple charge state heavy ion sources have been developed at several laboratories, which are primarily restricted to four types of ion sources: Electron Cyclotron Resonance Ion Source (ECRIS), Electron Beam Ion Source (EBIS), Metal Vapor Vacuum Arc ion Source (MEVVA), and Laser Ion Source (LIS). Pulsed highly charged heavy ion beams have been produced with an ECRIS with the afterglow mode at GSI [1], and CERN; EBISs invented by Donets in 1965 [2], have a very complicated system for beam control, and consequently they have huge dimensions compared to other ion sources; The beam intensity produced with a MEVVA ion source [3, 4] can reach to the order of mA, but the charge states are relatively low. In contrast, LIS can easily produce the high charge state and high intensity ion beams (several ~ mA), especially the refractory metallic ion beams, and what is more, the system only consists of simply instruments.

The collaboration between CERN, ITEP-Moscow, and TRINITI-Troitsk was devoted to developing a LIS for LHC, in which a $CO_2$ laser (100 J, 15-30 ns) was designed operate at a repetition rate of 1Hz. With this laser, a few emA peak current of $Pb^{27+}$ beam was achieved, and the total extracted current was about 20 emA [5 and 2]. Okamura proposed an idea of direct injection of the laser produced plasma into a RFQ cavity, or so-called direct plasma injection scheme (DPIS) [6], with which the maximum currents of 60mA of $C^{4+}$ [7] and 17mA of $C^{6+}$ [8] were obtained. The PERUN at the Institute of Physics in Prague (1315 nm, 50 J, 350-800 ps), with the laser intensity up to $10^{16}$ $W/cm^2$, was used to generate highly charged various elements: Al, Co, Ni, Cu, Ag, Sn, Ta, W, Pt, Au, Pb, Bi, and the maximum charge state achieved was up to 50+ [9].

Since 2007, the Institute of Modern Physics (IMP) started to develop laser ion sources to produce highly charge ion beams. Due to the capability of LIS to produce high-brightness ion beam (with current intensity of 10-100 emA and pulse duration of 1-10 μs), single-turn and single-pulse injection mode for a synchrotron could be realized. Therefore, LIS has been considered as a candidate ion source for a planned project, the High Intensity heavy ion Accelerator Facility (HIAF). For the application of LIS in accelerators and DPIS, the most important issue is the properties of laser produced plasma, which is the main topic of this paper.

# 2 LIS principle

In a laser ion source, a short pulsed, high power laser beam is focused to a small spot on a solid target leading to localize heating and subsequent evaporating. Electrons in the gas evaporated from the target absorb the laser irradiation via inverse Bremsstrahlung process and then turn into kinetic energy up to several hundreds of electron voltage. When the laser plasma occurs, the heavy ions in the plasma are pumped up by electron-ion collisions; ionization process is step by step, and then the laser plasma rapidly expands normally to the target surface within a small emitting angle. The temperature of the plasma and the consequent final ion charge states distribution strongly depend on the laser power density on target surface.

The cutoff density of the laser plasma is higher than that of other kinds of plasmas of conventional ion sources, such as ECR plasma. For instance, the $n_{cutoff}$ made by Nd:YAG laser ( λ= 1064 nm ) is $0.98 \times 10^{21}$ $cm^{-3}$,while $n_{cutoff}$ of an 24GHz ECR ion source is $7.16 \times 10^{12}$ $cm^{-3}$. The advantage of the high density plasma is available to produce many ions. The expression of plasma cutoff density is shown below:

$$n_{cutoff} = \frac{\varepsilon_0 m_e \omega^2}{e^2} = \frac{1.11 \times 10^{13}}{\lambda^2 [cm]} [cm^{-3}] \qquad (1) .$$

Where $e_0$: permittivity of free space, $m_e$: electron mass, $w$: angular frequency.

The electron temperature of the plasma which is determined by the Inverse Bremsstrahlung process is roughly as follows:

$$T_e \propto (I\lambda^2)^{2/3} \quad , \qquad (2)$$

where $T_e$: electron temperature, $I$: laser intensity [W/cm$^2$], $\lambda$: laser wavelength [$\mu$ m].

The inverse Bremsstrahlung process refers to the process in which an electron absorbs radiation as it scatters in the Coulomb field of an ion. The absorption coefficient is

$$\alpha = \frac{n_e e^2 \gamma_c}{\varepsilon_0 m_e c (\omega^2 + \gamma_c^2)} \quad , (3)$$

where $m_e$: the mass of electron, $n_e$: electron density, $\gamma_c$: electron-atomic collision frequency.

The layout of the experimental test bench of the LIS is shown in Fig.1 and the main parameters are listed in Table 1. The laser beam was transited for several meters in air by several high reflection (HR) flat mirrors, and then went into the vacuum target chamber through a K9 glass window. The target chamber and the diagnostic equipment were evacuated to about 10$^{-4}$ Pa. To avoid the damage to the laser generator optical-elements, a faraday isolator was used between the exit of the laser and the first HR mirror to eliminate the negative effects of optical feedback, which allows laser to pass unimpeded in the forward direction, while strongly attenuated in the backward direction.

## 3 Experimental setup

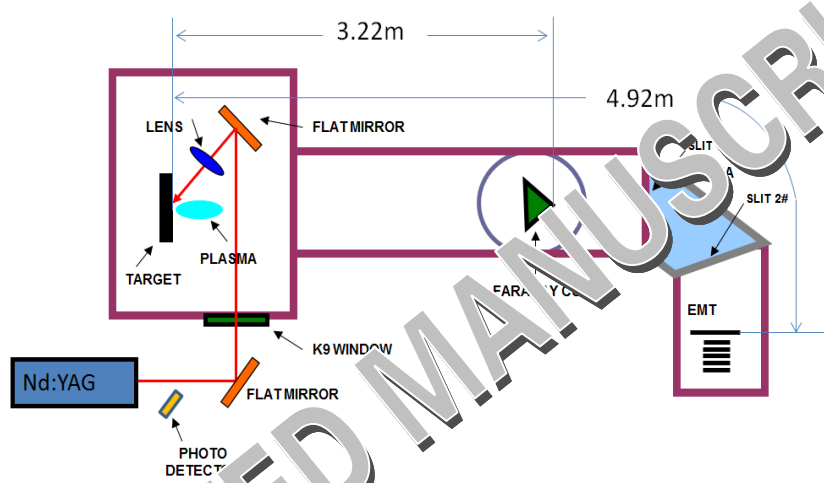

Fig.1. (color      ) L      f the laser plasma ion source experimental beam line setup.

Table 1.    The main parameter      Nd:YAG laser of
                                    t      s

| specification | value |
|---|---|
| laser type | Nd:YAG |
| wavelength/nm | 1064 |
| $E_{max}$/J | 3 |
| $E_{max}$(after faraday isolator)/J | 2.419 |
| pulse width/ns | 8-10 |
| Laser output diameter/mm | 15 |
| Repetition/Hz | 1 |
| Focal length/mm | 100 |
| Divergence/mrad | <1 |
| Incident angle | ~45$^o$ |

The laser beam was focused on the target by a convex lens held by a precise three dimensional manipulator consisting of three actuating motors. The minimum step of the 3-D manipulator is 0.01mm which can provide a fresh target surface after every laser shot. The focal length of the lens is 100 mm and the incident angle is about 45$^o$. The dimension of the target is 50 mm × 50 mm with 1mm thickness.

As mentioned in formula (2), the electron temperature $T_e$ is proportional to $(I\lambda^2)^{2/3}$, and the properties of the laser plasma strongly depends on the $T_e$. The laser intensity $I$ [W/cm$^2$] is given as follows:

$$I = \frac{E_{eff}}{\tau\pi(\frac{d_{eff}}{2})^2} \quad , \quad (4)$$

where $E_{eff}$: the effective energy of the laser; $\tau$: the laser pulse width; $d_{eff}$: the effective laser focal spot size on the target. As laser intensity being changed, the average charge state in the laser plasma will changed accordingly. When the laser generator is chosen, the $E_{eff}$ and $\tau$ are set. The laser intensity is then only influenced by $d_{eff}$. The $d_{eff}$ is subject to both the roughness of target surface and the location of the target at the laser beam waist. Furthermore, the highly charged ions expand in the normal direction of the target surface, and the jet angle is about $20^{o}$. The emission distribution is approximated by the superposition of a cosine and a $\cos^n$ fit function [10,11]. The good planeness surface of the target is favorable to the repeatability of the laser produced plasma and, to the adjustment of the target surface normal direction to match the beam line axis as well.

The photo detector can provide the trigger signal, which is generated by the diffused light from the first HR mirror. This trigger signal is used as the start time of the detector system. The distances from the target surface to the Faraday Cup (FC) and EMT detector are 1.92 m and 4.92 m, respectively. Those normal parameters can be used to calculate the ion pulse duration time $\tau_s$ and ion beam current density $j_s$ with the experimental scaling law [12]:

$$\tau_s \cong \tau_0(\frac{L_s}{L_0}) \quad j_s \cong j_0(\frac{L_s}{L_0})^3 \quad , \quad (5)$$

$L_0$: Measured position; $L_s$: Desired position; $\tau_0$: Pulse width time at $L_0$; $\tau_s$ Pulse width time at $L_s$; $j_0$: Current density at $L_0$; $j_s$: Current density at $L_s$.

The EIA is available to distinguish ion charge states, only the ions with a given energy-to-charge state ratio, $E/z$, can pass though the EIA [13].

$$\frac{E}{z} = \frac{eU}{[2\ln(\frac{R_2}{R_1})]} \quad , \quad (6)$$

where, $R_2$: outer plate radius; $R_1$: inner plate radius. In our case, $R_2$=505mm, $R_1$=495mm.

The main parameters of the EIA are listed in Table 2. To avoid the machining deformation, the special electrode was designed, shown in Fig.2.

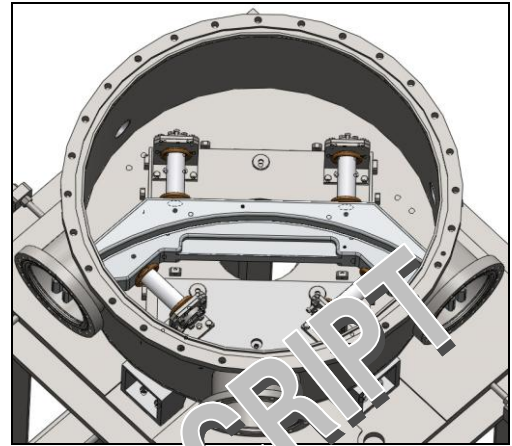

Fig.2. (color online) the field graph picture of EIA and vacuum chamber

Table 2 The mainly parameters of the EIA.

| name | value |
| --- | --- |
| Plate shape | Cylindrical |
| Plate material | Aluminum |
| bending angle | $90^{o}$ |
| $U_{max}$/kV | $\pm 40$ |
| $(E/z)_{max}$/eV | $\sim 10^6$ |
| $R_0$/mm | 500 |
| Gap/mm | 10 |
| Electrode height/mm | 60 |
| Vacuum in tank/Pa | $10^{-5}$ |

Two slits are placed at the entrance and the exit of the EIA. The width range is ~5 mm from the center. The slits are used to improve the resolution of the EIA. The voltage applied to the electrode of the EIA can reach up to $\pm 40$ kV. At the downstream, the EMT detector was placed. The EMT detector has high gain yet low dark current and is therefore useful to detect very small amount of charged particles and particles with low energy. Using the EMT detector with EIA, the spectrum of the flight time of different charge state ions can be achieved from the oscilloscope that connects with the EMT detector.

## 4 Laser plasma production

Because the focal position is crucial to the production of the high charge state ions as mentioned above, two tasks should be done before measuring laser plasma: 1). the target surface flatness needs to be measured; 2). optimize of the target position.

To measure the target surface flatness, the KEYENCE High-speed, High-accuracy CCD laser and displacement sensor was used. Fig.3 shows the typical data plot of the target surface. In general, the flatness of the target surface needs to be better than 0.05 mm, which can benefit for the good repetitiveness of the laser plasma condition. As shown in Fig.3, the surface flatness of this target was about 0.0466 mm. With the three dimensional manipulator and the FC signals, the optimization of the focal position of the target is easily achieved.

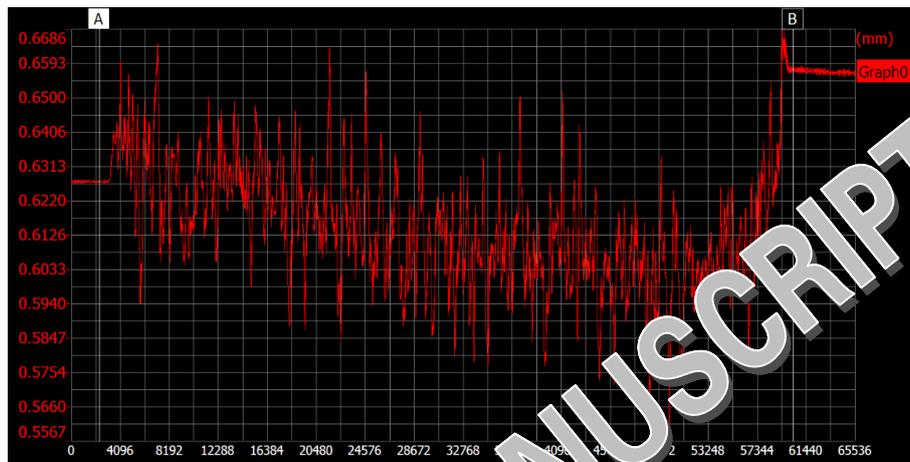

Fig.3. Flatness data of the target surface

Abscissa is the data, the measurement time

As shown in Fig.4 (laser beam conditions: E = 2.297 J, 1064 nm, f = 100 mm), the highest peak current and lowest peak time can be easily achieved, which were the obvious evidence of the best focal position of the laser shot. When the target position was chosen, the voltage of EIA detector would be scanned to get ion spectra.

detector. The peaks in the spectrum could be attributed to ions with different energy-to-charge ratio. Due to the principle of the EMT detector, the signal comes from the secondary electron, so the negative signal is shown. $H^+$ came from the contamination of the residual gas in the target chamber or residual water on the target surface when the target was installed. The base line of the spectrum was high because the first slit of EIA was opened widely.

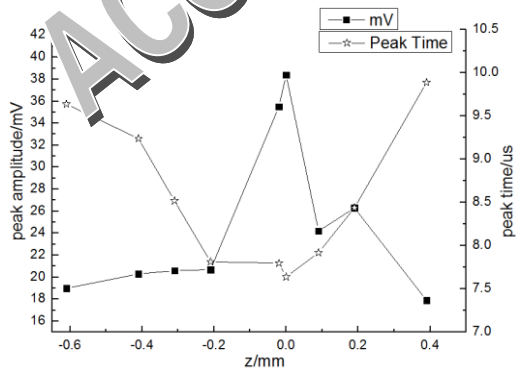

Fig. 4. Peak time and beam current of FC with different longitudinal positions of the target

Fig.5 shows the typical signal from the EMT

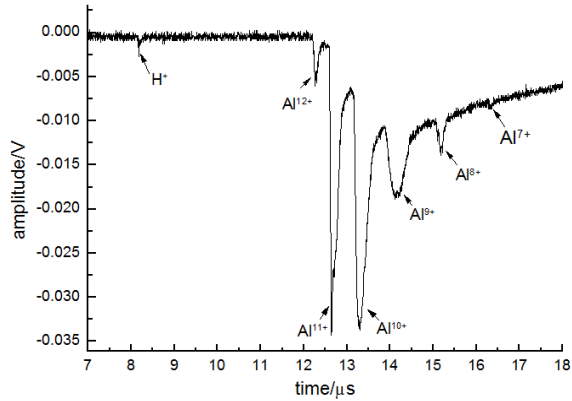

Fig. 5. Typical spectrum from the EMT detector.

Fig.6 and Fig.7 show results for the aluminum material, Fig.8 and Fig.9 for the lead material, respectively. The charge states of aluminum and lead ranged from $Al^{3+}$ to $Al^{12+}$ and from $Pb^{1+}$ to $Pb^{7+}$, respectively. From the figs.8, one can see that the higher charged state ions concentrate at the early stage of the spectrum, while lower charged state ions extend to a long tail. The highest yield ions are $Al^{10+}$ (23.91 %) and $Pb^{4+}$ (20.95 %), respectively. The measured total ion beam currents at the FC position are 0.54 mA for Al and 0.48 mA for Pb.

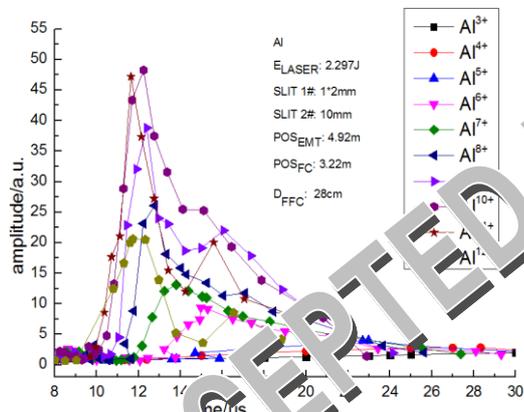

Fig.6. (color online) Charge state distribution obtained with Al target.

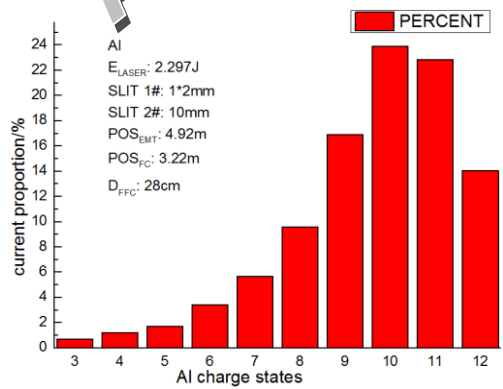

Fig.9. Percentage of different charge state of the Pb beam.

Fig.7. Percentage of different charge states obtained with Al target.

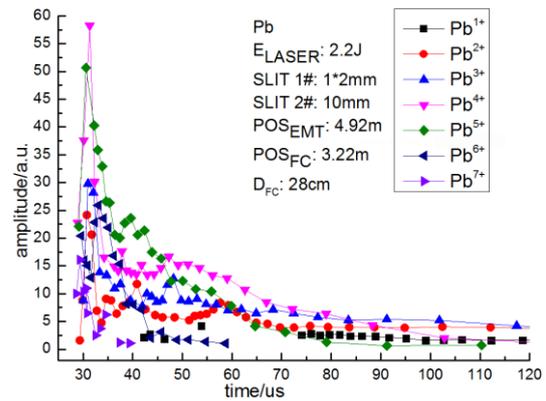

Fig.8. (color online) Charge states distribution of Pb beam.

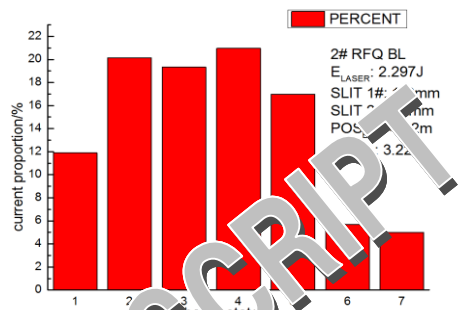

Fig.9. Percentage of different charge state of the Pb beam.

## 5 Discussion

As presented above, in our laser plasma production experiment, very highly charged Al ions have been achieved; while for Pb plasma, the highest charge state was $Pb^{7+}$, whose ionization potential is much lower than that of $Al^{12+}$. The possible reason for this great difference might be: 1) The laser power density is not high enough. In our case it is about $10^{12}$ W/cm$^2$, which is not suitable for high charge states of heavy element material, like Pb. 2) During the early stage of laser plasma expansion, recombination [14,15] of highly charged ions might occurs, which would depress the high charge state yield. From the spectra presented in Fig.6 and Fig.8, the peak velocities of Pb and Al ion are estimated to be $1.03 \times 10^5$ and $4.06 \times 10^5$ m/s, respectively, which makes the transit time of Al ions through the recombination zone is much shorter than that of Pb ions and finally leads to the great difference in the charge state distribution of the two species. This effect needs to be confirmed by further investigations. .

# 6 Conclusion

At IMP, we have built a beam line for the laser plasma research. The aluminum and lead foil targets have been tested; the high charge state ion beams of $Al^{12+}$ and $Pb^{7+}$ has been achieved.

In the future experiment, we intend to use different incident angle and laser energy to study their relationships to the ion beam average charge state; and still plan to use different pure material or some compound material targets to test the production of highly charged state ions.

# Reference

1 Schulte H, Bossler J, S. Schennach et al. Rev. Sci. Instrum., 1994, **65**:1081

2 Donets E D, Evgeni D. Rev. Sci. Instrum.,1998, **69**:614

3 Brown I G, Rev. Sci. Instrum. 1994, **65**: 3061

4 Brown I G and Oks E, IEEE Trans. Plasma Sci., 2005, **33**: 1931

5 Alessi J G, Proceedings of LINAC 2004, http://www.bnl.gov/isd/documents/28625.pdf

6 Okamura M, Takeuchi T, et al, in Proceedings of the European Particle Accelerator Conference on Design Study of RFQ Linac for Laser Ion Source (European Physical Society Interdivisional Group on Accelerators, Vienna, 2000), 2000, 848–850.

7 Okamura M, Kashiwagi Hirotsugu, et al, Rev. Sci, Instrum. 2006, **77**:03B303

8 Kashiwagi H, Fukuda M, et al, Rev. Sci. Instrum., 2006, **77**:03B305

9 Laska L, Jungwirth K, et al., Rev. Sci. Instrum., 2004, **75**:1546

10 Thum-Jager Andrea and Rohr Klaus. J. Phys. D: Appl. Phys., 1999, **32**: 2827

11 Laska L, Krasa J, Pfeifer M, et al, Rev. Sci, Instrum., 2002, **73**: 654

12 Sharkov B Yu and Kondrashev S, in Proceedings of the European Particle Accelerator Conference on Matching of the Intense Laser Ion Source to the RFQ Accelerators (Taylor & Francis, Sitges, 1996), 1996, 1550–1552

13 Woryna E, Parys P, et al., Laser and Particle Beams, 1996. **14**:

14 Roudskoy V I, Laser and particle beams, 1996, **14**:369

15 Kondrashev S, Kanesue T, et al., J. Appl. Phys., 2006, **90**: 3501